\documentclass[baaa]{baaa}

\usepackage[pdftex]{hyperref}
\usepackage{subfigure}
\usepackage{natbib}
\usepackage{helvet,soul}
\usepackage[font=small]{caption}
\contriblanguage{1}

\contribtype{1}

\thematicarea{99}

\received{\ldots}
\accepted{\ldots}

\title{The new era of Lyman alpha emitters (LAEs): \\ Typical star formation histories of LAEs in the ILLUSTRIS simulation}

\titlerunning{The new era of Lyman alpha emitters}

\author{
I. Laferte-Urrutia\inst{1}, P. Troncoso-Iribarren\inst{1}, Alex Vera-Casanova\inst{1, 2}, C. Artale\inst{3}, L. Guaita\inst{2}, E. Gawiser\inst{4,5}, J. Magaña \inst{1},  C. Vega-Martínez \inst{1}, K. Lee\inst{6}, N. Firestone\inst{4} \& P. Layana-Astudillo\inst{1}
}

\authorrunning{I. Laferte-Urrutia et al.}

\contact{ivonne.laferte@alumnos.ucentral.cl}

\institute{
Facultad de Ingeniería y Arquitectura, Universidad Central de Chile, Chile
\and
Departamento de Astronomía, Universidad de La Serena, Raúl Bitrán N° 1305, La Serena Chile.
\and
Instituto de Astrofísica, Universidad Andrés Bello, Chile
\and
Department of Physics and Astronomy, Rutgers, the State University of New Jersey, Piscataway, NJ 08854, USA
\and
School of Natural Sciences, Institute for Advanced Study, Princeton, NJ 08540, USA
\and
Department of Physics and Astronomy, Purdue University
}

\resumen{%El muestreo ODIN ha detectado miles de emisores Lyman-$\alpha$ (LAEs) en siete campos del cielo, cubriendo un área total de casi 100 grados cuadrados \citep{Lee_2024}.%;\citep{firestone2024odinimprovednarrowbandlyalpha} \citet{firestone2025odinstarformationhistories} modelaron las historias de formación estelar (SFH) de 73 LAEs muestreadas entre 0.4 y 10 $\mathrm{\mu m}$, identificando tres tipos de SFH típicas a z = 2.4, 3.1 y 4.5. A z = 2.4, el 77 \% de las LAEs presenta una SFH donde la mayor actividad ocurre en el momento de observación, mientras que un 16 \% muestra múltiples bursts dominantes y solo un 7 \% no exhibe bursts dominantes.
Este trabajo busca comprender la naturaleza de las galaxias emisoras de Lyman-$\alpha$ en un contexto cosmológico analizando las historias de formación estelar en la simulación ILLUSTRIS-TNG100, aplicando el criterio de selección reciente. La muestra a z = 2.0 incluye 6051 galaxias, clasificadas en cuatro clases (35 \%, 33 \%, 21 \% y 11 \%) mediante \textsc{KMeans}, un método de machine learning no supervisado. La primera clase reproduce la historia de formación estelar típica caracterizada por una formación estelar más intensa al momento de observación. Las restantes presentan historias atípicas con brotes estelares a 0.3, 0.7 y 1.3 $\mathrm{Gyr}$ antes del momento de observación. La primera clase corresponde a las galaxias de menor masa, luminosidad Lyman-$\alpha$ y tasa de formación estelar.
Se concluye que la definición clásica de Lyman alpha emitters de baja masa, poco polvo y un único brote sigue siendo la más representativa (35 \% del total de la muestra), pero existen otras clases que abarcan el 65\% de la muestra cosmológica total.}

\abstract{%The ODIN survey has detected thousands of Lyman-$\alpha$ emitters (LAEs) in seven fields of the sky, covering a total area of nearly 100 square degrees \citep{Lee_2024}.
This work seeks to understand the nature of Lyman-$\alpha$ emitting galaxies in a cosmological context by analyzing their star-formation histories in the ILLUSTRIS-TNG100 simulation, applying a recent selection criteria. The sample at z = 2.0 includes 6051 Lyman alpha emitters, classified into four classes (35\%, 33\%, 21\%, and 11\%) using \textsc{KMeans}, an unsupervised machine learning clustering method. The first class reproduces the typical star-formation history, characterized by the most intense star formation at the time of observation. The remaining classes exhibit atypical star-formation histories, with bursts at 0.3, 0.7, and 1.3 $\mathrm{Gyr}$ before the time of observation. The first class corresponds to galaxies with lower mass, Lyman-$\alpha$ luminosity, and total star-formation rate. We concluded that the classic definition of Lyman-$\alpha$ emitting galaxies—low mass, low dust, and a single burst—remains the most representative (35\% of the total sample), but there are other classes that encompass 65\% of the total cosmological sample.}
\keywords{galaxies: high-redshift --- galaxies: evolution --- methods: observational --- methods: numerical}

\begin{document}

\maketitle
\section{Introduction}\label{S_intro}

The large-scale structure (LSS) of the Universe provides crucial information about the formation and distribution of galaxies and matter throughout cosmic time. To study these structures, we use Lyman alpha emitters (LAEs) as tracers. The intense emission of the Ly$\alpha$ spectral line serves as a probe of ultraviolet (UV) light originating in star-forming regions, active galactic nuclei, and other energetic processes. LAEs are typically young galaxies with low stellar mass, metallicity, and dust content \citep{2006ApJ...642L..13G}, allowing more Ly$\alpha$ photons to escape in comparison to older and dustier systems. 
According to observational results summarized in the annual review of \citet{2020ARA&A..58..617O}, LAEs at redshift $z \geq 2$ are characterized by low UV luminosities ($M_{\mathrm{UV}} \geq -20$), rest-frame equivalent width (REW $\geq 20$~\AA), and stellar masses of $\log_{10}(M_{\star}/M_{\odot}) \approx 8$–$9$. Yet, more recent results suggest that the star formation histories (SFHs) of LAEs may be significantly more diverse and complex. 
%In particular, the study by \citet{firestone2025odinstarformationhistories} proposes the existence of three distinct SFHs for LAEs based on a sample of 39 LAEs at z=2.4.
\citet{firestone2025odinstarformationhistories} modeled the SFH of 73 LAEs sampled between 0.4 and 10 $\mathrm{\mu m}$, identifying three types of typical SFHs at z = 2.4, 3.1, and 4.5. At z = 2.4, 77\% of LAEs have a SFH where the highest stellar activity occurs at the time of observation, 16\% show multiple dominant bursts, and 7\% show no dominant bursts. 

In this work, we analyze the SFHs in a cosmological context, specifically from the IllustrisTNG-100 simulation \citep{2015A&C....13...12N}.
By adopting the criteria described by \citet{2025A&A...698A.280A}, we select the LAEs at $z = 2.0$, with stellar masses in the range $8.0 \leq \log_{10}(M_{\star}/M_{\odot}) \leq 11.5$. This range allows for direct comparison with previous observational studies \citep{firestone2025odinstarformationhistories}.
It results in a sample of 6,070 LAE galaxies identified by their subhalo IDs, along with their corresponding physical properties, such as stellar mass, star formation rate (SFR), luminosity, and REW magnitudes. %This approach allows for a robust and comparative study of the properties of LAEs across different cosmic epochs.
%The SFH of a galaxy can be reconstructed by following the main progenitor branch of each subhalo through the simulation’s snapshots. 
We follow the SFH for halos that exhibit characteristics of LAEs in  IllustrisTNG-100 across different cosmic epochs. 

\section{Methodology}

\subsection{IllustrisTNG hidrodynamical simulation}
The IllustrisTNG project \citet{2015A&C....13...12N} is a suite of large-scale, magneto-hydrodynamical cosmological simulations designed to model the formation and evolution of galaxies across cosmic time within the $\Lambda$CDM framework. The project comprises three main simulation volumes: TNG50, TNG100, and TNG300, with comoving box sizes of approximately 50, 100, and 300 $\mathrm{Mpc}$, respectively.
The simulations were developed using the moving-mesh \textsc{Code Arepo} \citep{2010MNRAS.401..791S}, which coupled the evolution of dark matter (DM), gas, stars, and supermassive black holes from a high redshift ($z \approx 127$) to the present day ($z = 0$). For this work, we choose the TNG100-1 box. The resolution of mass particles of DM \textbf{\( m_{\mathrm{DM}} \approx 5.1 \times 10^{6}~\mathrm{M_{\odot} h^{-1}}\)} and \textbf{\( m_{\mathrm{b}} \approx 9.4 \times 10^{5}~\mathrm{M_{\odot} h^{-1}} \)} for baryons, while the temporal resolution at \( z = 2.0 \) is approximately \( 1.5~\mathrm{Myr} \).
% which models a comoving cube of \( (110.7\ \mathrm{Mpc})^3 \), 
%It is large enough to include a statistically representative population of galaxies and groups, and it maintains adequate spatial and mass resolution to study the internal structure of individual galaxies. The numerical resolution corresponds to particle masses of \( m_{\mathrm{DM}} \approx 5.1 \times 10^{6}~M_{\odot}/h \) for dark matter and \( m_{\mathrm{b}} \approx 9.4 \times 10^{5}~M_{\odot}/h \) for baryons, while the temporal resolution at \( z = 2.0 \) is approximately \( 1.5~\mathrm{Myr} \).
%This model provide a diverse population of galaxies and groups, and it maintains adequate spatial and mass resolution to study the internal structure of individual galaxies.
%The particle resolution used by the dark matter (DM) and baryonic matter (b) simulation is $m_{\mathrm{DM}} \approx 5.1 \times 10^{6}~M_{\odot}/h$, $m_{\mathrm{b}} \approx 9.4 \times 10^{5}~M_{\odot}/h$ and the time-resolution at $z=2.0$ is $\mathrm{\sim1.5 \, Myr}$
The simulation adopts the cosmology of \citet{refId0}, with the following parameters:
\(
\Omega_{\mathrm{m}} = 0.3089, \quad
\Omega_{\mathrm{b}} = 0.0486, \quad
\Omega_{\Lambda} = 0.6911, \quad
h = 0.6774.
\)
%\subsection{Luminosity function}

To assign Ly$\alpha$ emission line luminosities ($L_{\mathrm{Ly}\alpha}$) and REW to galaxies, we adopt the empirical model introduced by \citet{2012MNRAS.419.3181D}.
Specifically, the $L_{\mathrm{Ly}\alpha}$
%Ly$\alpha$ luminosity 
is then obtained from the UV luminosity according to equation 4 of \citet{2025A&A...698A.280A}. This criterion is also used by \citet{Im_2024} to test the validity of LAEs 
as tracers of the LSS %large-scale structures 
of DM halos at $z=2-4$.

%Although these works calibrated their parameters to reproduce Ly$\alpha$ luminosity functions at $z \sim 2.4$, $3.1$, and $4.5$ , here we apply the same formalism to our sample at $z = 2.0$. The Ly$\alpha$ luminosity is then obtained from the UV luminosity using:

% \begin{equation} 
% \label{eq 1}
% L_{\mathrm{Ly}\alpha} = \mathrm{REW} \times L_{\mathrm{UV}}\times \frac{\nu_{\alpha}}{\lambda_{\alpha}}\bigg(\frac{\lambda_{\alpha}}{\lambda_{\mathrm{UV}}}\bigg)^{-\beta-2}  
% \end{equation} 
% Where $\nu_{\mathrm{\alpha}}$ and $\lambda_{\mathrm{\alpha}}$ are the frequency and wavelength of the Ly$\alpha$ emission line, respectively. The $L_{\mathrm{UV}}$ represents the UV
% luminosity, and $\lambda_{\mathrm{UV}}$ is the wavelength of the UV emission.
% Here, we assume that the UV continuum of each galaxy follows a power law with a slope of $\beta$. \\

%\subsection{Physical properties of LAEs in the Illustris TNG simulation}
%Galaxies are identified using the \textsc{SubhaloID} parameter, which allows access to individual properties stored at \textsc{Snapshots 33} \footnote{\url{{https://www.illustris-project.org/data/docs/specifications/}}}, 
Galaxies are identified using the \textsc{SubhaloID} parameter and follow the properties stored at \textsc{Snapshots 33} \footnote{\url{{https://www.illustris-project.org/data/docs/specifications/}}}, equivalent to a redshift $z \sim 2.0$. %The galaxies and halos we see in a snapshot are not static objects; they are born, grow, merge, and disappear over time.
We used the \textsc{Merger trees} to follow the the temporal relationships between subhalos at different snapshots,
%\textsc{Merger trees} were used as a data structure that records the temporal relationships between subhalos that exist in different snapshots,
allowing track the history of each LAE. We traced from birth in the early universe to the observed time at $z=2$.
%using the \textsc{Sublink} algorithm.
The \textsc{SubhaloSFR} variable, which represents the instantaneous SFR of the galaxy at each epoch in units of $\mathrm{M_\odot yr^{-1}}$, was extracted throughout the tree. An additional filter based on the stellar mass ($M_{\star}$) of the galaxies was applied, considering those with masses in the range $8.0 \leq \log_{10}(M_{\star}/M_{\odot}) \leq 11.5$ to focus on intermediate galaxies that are representative of the observed populations. The selection was based on the variable stellar mass particles, converted to physical units using cosmological factors already described. \\
%Of the 6070 galaxies identified by \citet{2025A&A...698A.280A}, the SFHs of 6051 were recovered; the remaining 19 lacked valid SFR data, probably due to dynamic complexity or simulation limitations (e.g., insufficient stellar particles, DM, or gas).

%This selection was made by accessing the \textsc{SubhaloMassType} variable, which was converted to physical units using the appropriate cosmological factors. \\
%Finally, of the $6,070$ galaxies initially identified by \citet{2025A&A...698A.280A}, the SFHs were recovered for 6,051 galaxies. The remaining 19 galaxies were discarded from the analysis due to the absence of valid information on their SFR. They were possibly affected by complex dynamic processes or numerical limitations of the simulation; e.g., they did not reach the minimum number of stellar particles, DM, or gas required to form a galaxy. 

%\textcolor{red}{citar andrews}
% Of the $6.070$ galaxies initially identified, we were able to recover the SFH of $6.051$ galaxies, including physical properties such as spatial position, stellar mass, star formation rate (SFR), UV luminosity, rest-frame equivalent width (REW), etc. The remaining 19 galaxies
% were discarded from the analysis due to the absence of valid information on their star formation rates, possibly affected by complex dynamic processes or
% by numerical limitations of the simulation. 

\subsection{Star formation histories of LAEs}

%The galaxies and halos we see in a snapshot are not static objects; they are born, grow, merge, or disappear over time. \texttt{Merger trees} were used, a data structure that records the temporal relationships between subhalos or galaxies that exist in different snapshots, allowing the complete evolutionary history of a galaxy to be traced from its birth in the early universe to the present day using the \texttt{Sublink} algorithm. The \texttt{SubhaloSFR} variable was extracted from the tree, representing the instantaneous star formation rate of the galaxy at each epoch (in units of $\mathrm{M_\odot \, yr}^{-1}$). \\

The SFR describes the amount of stellar mass formed per unit time in a galaxy, expressed in $(M_{\odot} \,\mathrm{yr^{-1}})$. By plotting the temporal evolution of the SFR as a function of the lookback time (LBT), it becomes possible to visualize the different episodes of SF experienced by a galaxy, including rises, declines, or a single dominant burst of stellar activity. This analysis is essential to understanding the processes of stellar activation, quenching, or rejuvenation throughout a galaxy's lifetime.
In cosmology, the LBT represents the interval of time elapsed between the moment when light was emitted by a distant source and the present moment when that light is observed. Mathematically, it is defined as the difference between the current age of the Universe and the age of the Universe when the light was emitted:

\begin{equation}
    \mathrm{LBT(z)} = t(z = 0) - t(z),
    \label{eq:LBT_general}
\end{equation}
where $t(z=0)$ is the current age of the Universe ($\sim 13.8~\mathrm{Gyr}$), and $t(z)$ is the age of the Universe at the corresponding redshift $z$ of the emitting source. \\
In this work, we analyze LAE galaxies located at an average redshift of $z \sim 2.0$, which corresponds to a LBT $ \approx 10~\mathrm{Gyr}$. %In other words, we observe these galaxies when the Universe was about $72\%$ of its current age. 
%To consistently study their temporal evolution, it is necessary to redefine the LBT in the reference frame of each galaxy, taking as the zero point the moment at which the galaxy is observed ($z = 2.0$).

%\textcolor{red}{aqui idente 2da ecuacion}
% \begin{equation}
%     LBT_i = LBT(z_i) - LBT(z = 2.0)
%     \label{eq:LBT_relative}
% \end{equation}
% where $z_i$ corresponds to the redshift of snapshot $i$, and $LBT(z=2.0)$ represents the observational time of the galaxy in the simulation.

The estimation of $\mathrm{LBT(z)}$ is performed using the \textsc{astropy.cosmology} library, assuming a flat $\Lambda$CDM cosmology consistent with the parameters adopted in the IllustrisTNG simulation. The analyzed snapshots span the redshift range $ z = 20.046 $ to $ z = 2.002 $, corresponding to snapshots 99 to 33 and covering a temporal interval of $\sim 3.0\,\mathrm{Gyr}$.
%Since galaxies display different absolute SFR values, it is necessary to normalize each galaxy’s SFH to compare them on a common scale.
To compare galaxies with varying absolute SFRs, each SFH is normalized using its own maximum and minimum SFR values. Galaxies were classified using a single parameter: the time of maximum SFR over their lifetime. For each galaxy, the SFHs were extracted and the SFR normalized to its peak along the main progenitor branch. After sorting snapshot redshifts, the peak normalized SFR defines $z_{\mathrm{peak}}$, the redshift of maximum SF. The normalized SFR is given by:

\begin{equation}
    \mathrm{SFR_{n,i} = \frac{SFR_i - SFR_{min}}{SFR_{max}}},
    \label{eq:SFR_normalized}
\end{equation}
where $\mathrm{SFR_{i}}$ is the SFR at snapshot $i$. $\mathrm{SFR_{max}}$ and $\mathrm{SFR_{min}}$ are the peak and minimum values over the galaxy's history.

%The classification was based on a single parameter called the time at which the maximum SFR occurs, considering the entire lifetime of the galaxy as the time interval.
%For each galaxy, this value is obtained by extracting the SFH, specifically the SFR normalized to its own maximum peak along its main progenitor branch.
%The redshifts of the snapshots are sorted, and the maximum value of the SFR is identified and used to normalize the SFR in each snapshot. The redshift at which the normalized SFR reaches its maximum in the trajectory of each subhalo corresponds to our $z_{peak}$. We define the variable ``normalized SFR'' as follows:

% \begin{equation}
%     SFR_{n,i} = \frac{SFR_i - SFR_{min}}{SFR_{max}},
%     \label{eq:SFR_normalized}
% \end{equation}

%where $SFR_i$ is the SFR at snapshot $i$, $n$ is the galaxy identification. $SFR_{max}$, and $SFR_{min}$ correspond to the maximum, minimum SFR values across the entire history of the galaxy, respectively.
%This normalization allows us to unambiguously identify the moment when the maximum SF burst occurs, $t_{\mathrm{peak}}$, corresponding to the snapshot where $SFR_{n,i}$ reaches its maximum value. This instant can also be expressed in terms of redshift ($z_{\mathrm{peak}}$) or relative LBT ($X_{\mathrm{peak}}$):

This normalization defines the peak SF epoch, ${t_{\mathrm{peak}}}$, as the snapshot where $\mathrm{SFR_{n,i}}$ is maximal. This moment can also be expressed as a redshift ${z_{\mathrm{peak}}}$ or relative LBT ${X_{\mathrm{peak}}}$. We define $X_{\mathrm{peak}}$ as the difference in LBT between the redshift of peak SF and the observation redshift \(z = 2.0\)

\begin{equation}
    X_{\mathrm{peak}} = \mathrm{LBT}(z = z_{\mathrm{peak}}) - \mathrm{LBT}(z = 2.0),
    \label{eq:Xpeak}
\end{equation}

\subsection{Machine learning method: KMeans} 
To classify the the LAEs, we normalized the SFHs and extracted the time of peak star formation ($t_{\mathrm{peak}}$), and applied the \textsc{KMeans} algorithm, an unsupervised method that partitions the data into $k$ groups, each defined by a centroid at similar $t_{\mathrm{peak}}$, representing the same SFH type. 
%evolutionary patterns in the SFHs of the LAE sample, we applied the \textsc{KMeans} clustering algorithm, an unsupervised method that partitions data into $k$ groups, grouping together galaxies with similar characteristics. Each cluster is defined by its centroid, representing the average SFH profile of its members.
%This approach reveals natural groupings, particularly variations in the timing of peak star formation ($t_{\mathrm{peak}}$). We explored multiple configurations ($k = 3,4,5$) to capture these differences. \\
%This allow, particularly variations in the timing of peak star formation ($t_{\mathrm{peak}}$). We explored multiple configurations ($k = 3,4,5$) to capture these differences. \\
This allows exploration of multiples configurations of types ($k = 3,4,5$) to obtain the best clasification. 
\textsc{KMeans} minimizes intra-cluster variance, yielding compact, well-separated clusters. Its simplicity, scalability, and interpretable centroids make it ideal for assigning representative SFH profiles to galaxy classes.
To evaluate clustering quality, we used a Silhouette score, which measures the compactness of each cluster and the distance between the clusters. Since \textsc{KMeans} initializes the centroids randomly, we performed multiple runs with different seeds to obtain the best fit. 
% The configuration with the highest \textsc{Silhouette score} for the different runs was selected.
%, and is defined as:
%\begin{equation}
%    s = \frac{b - a}{\max(a, b)}
%   \label{eq:silhouette}
%\end{equation}
%where $a$ is the mean distance between a galaxy and the other members of its own cluster (intra-cluster compactness), and $b$ is the mean distance to the nearest cluster to which it does not belong (inter-cluster separation).
%¿The Silhouette coefficient ranges from $-1$ to $+1$: values close to $+1$ indicate well-defined, compact clusters, values near $0$ suggest overlapping or boundary cases, and negative values imply misclassified galaxies. A high mean Silhouette score (typically $>0.6$) therefore denotes coherent and well-separated groups, validating the reliability of the KMeans classification applied to the SFHs.?
%\textcolor{blue}{Tambien mencionar que Kmean hace distribuciones de grupos basados en una semilla aleatoria, y es por eso que repetiste el experimento hartas veces, y por eso fue necesario usar una metrica que nos indico cual fue la mejor distribucion}
\section{Results}\label{sec:guia}
\begin{figure}[t]
%\centering
\includegraphics[width=0.96\columnwidth]{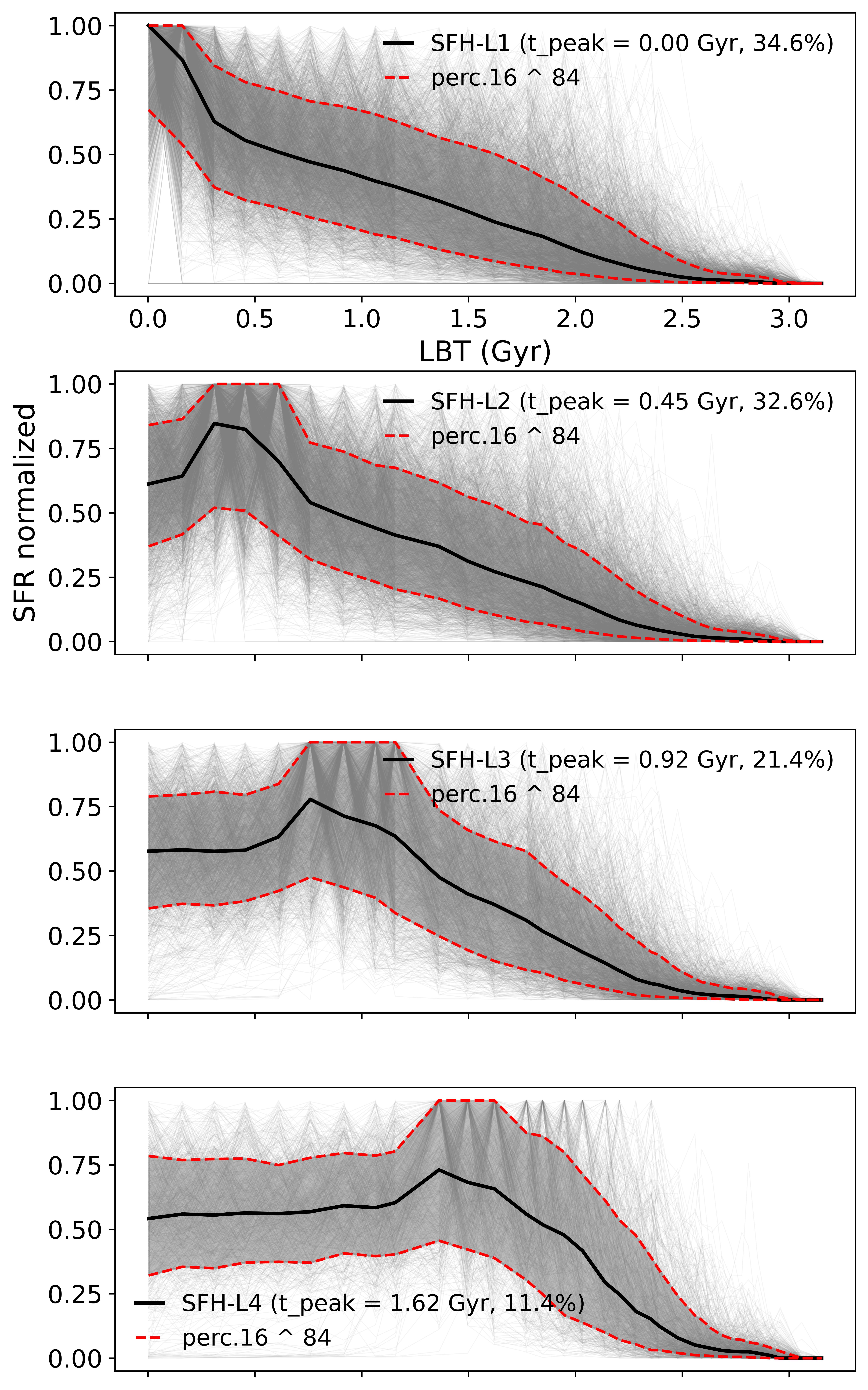}
\caption{SFH for $k=4$. The gray lines indicates the normalized SFR of each individual galaxy per class as a function of LBT, in $\mathrm{Gyr}$. The solid black line indicates the median of the normalized SFR. The dashed line represents the 84 and 16 percentiles are $1\pm \sigma$ above and below the mean, respectively. The label indicates the percentage of galaxies from the total sample, and the $t_{\mathrm{peak}}$ of each class.}
\label{Figura_vert}
\end{figure}

We present the results for each class obtained by \textsc{KMeans} , summarized in Table \ref{tab:kmeans_comparison_2option}. It presents a comparison of $t_{\mathrm{peak}}$, stellar mass, and SFR.
%Table \ref{tab:kmeans_comparison_2option} shows the results for each class obtained according to three different clustering groups, namely $k=3$, $k=4$, and $k=5$.
%It presents a comparison of the physical properties of each class; the stellar mass and SFR, ordered by increasing value of $t_{peak}$.
The best fit is obtained for $k = 4$. This indicates that the best classification is for four groups. These are labeled by SFH–L1 to SFH–L4. In Fig. \ref{Figura_vert}, we present the data distribution for each class for $t_{\mathrm{peak}}$ and M$_\star$. The gray lines demarcate the individual SFHs belonging to each class. The black line indicates the median of the normalized SFR, while the dashed-red lines indicate the 84th and 16th percentiles  $1\pm \sigma$, above and below the median, respectively.

%The y-axis shows the SFR normalized, while the x-axis shows the LBT, where $LBT = 0 \, \mathrm{[Gyr]}$ indicates the present time of the galaxy at $z=2.0$ and grows to the right, towards its past and birth. 
%In figure \ref{viol-tpeak}, the distribution of $t_{\mathrm{peak}}$, stellar mass is shown for the four classes ($k=4$), respectively. The black continuous line shows the median of the sample, while the dashed gray lines correspond to the 84th and 16th percentiles. 

\begin{figure}[!ht]
\centering
\includegraphics[width=0.77\columnwidth]{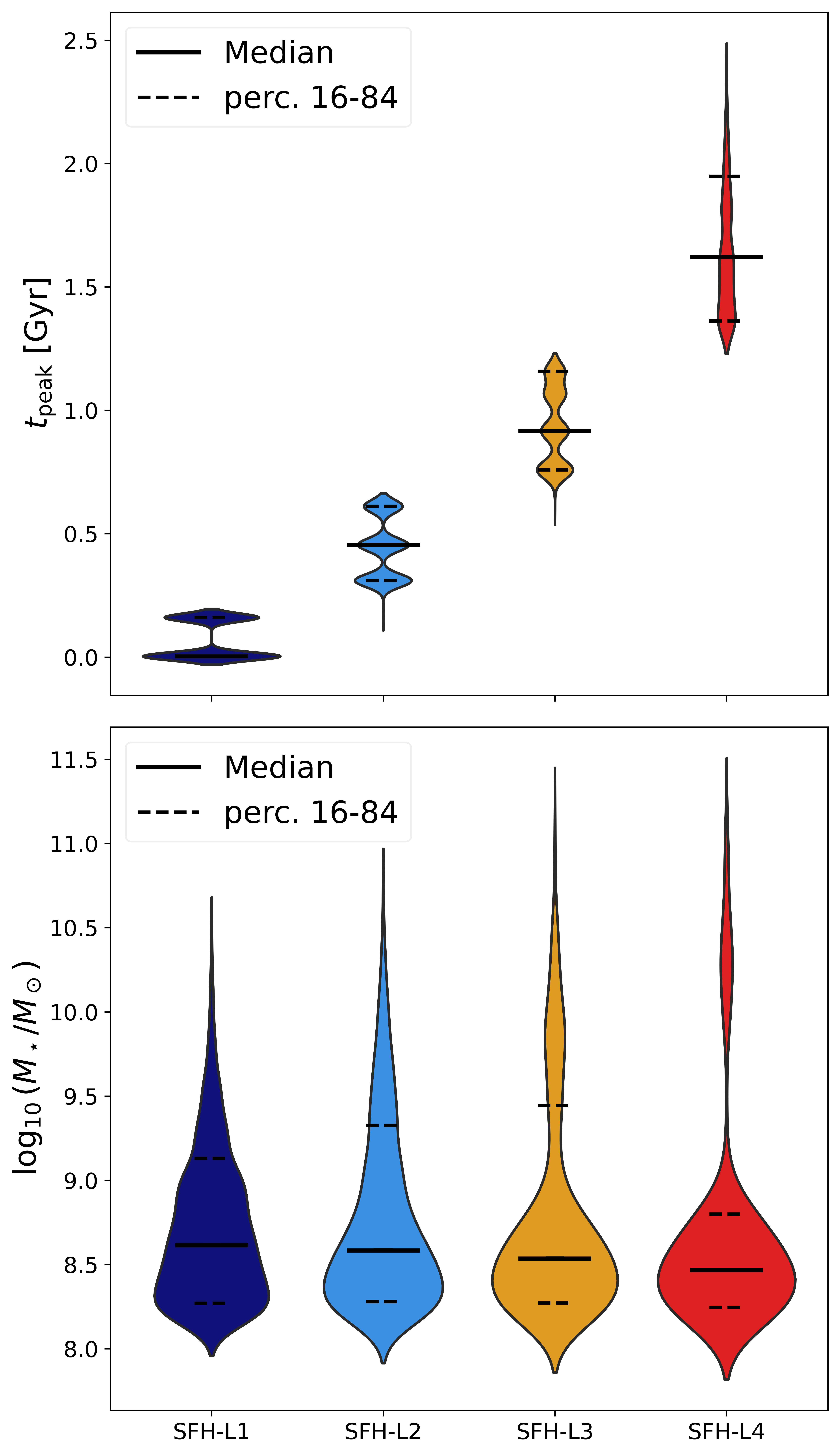}
\caption{Distribution of $t_{\mathrm{peak}}$ and $M_{\star}$ per SFH class, with $k=4$. The solid black line show the mean and dashed black lines correspond to the 84th and 16th percentiles.}
\label{viol-tpeak}
\end{figure}

\begin{table*}[t]
\centering
\scriptsize
%\footnotesize
\renewcommand{\arraystretch}{0.9}
\setlength{\tabcolsep}{3pt}
\caption{Comparison of mean properties of LAE clusters for different $k$ values.}
\label{tab:kmeans_comparison_2option}
%\begin{tabular}{@{}lcccl@{}|@{}lcccl@{}|@{}lcccl@{}}
\begin{tabular}{|lcccc|lcccc|lcccc|}
\hline\noalign{\smallskip}
\multicolumn{5}{c|}{$k = 3$, Silhouette = 0.64} &
\multicolumn{5}{c|}{$k = 4$, Silhouette = 0.65} &
\multicolumn{5}{c}{$k = 5$, Silhouette = 0.66} \\
\hline\hline\noalign{\smallskip}
Class  & \% & $t_{\mathrm{peak}}$ & $M_{\star}$ & SFR & Class & \% & $t_{\mathrm{peak}}$ & $M_{\star}$ & SFR & Class & \% & $t_{\mathrm{peak}}$ & $M_{\star}$ & SFR \\
~ & ~ & \scriptsize{[Gyr]} & \scriptsize{[$M_\odot$]} & \scriptsize{[$M_\odot$/yr]} &  
~ & ~ & \scriptsize{[Gyr]} & [\scriptsize{$M_\odot$]} & \scriptsize{[$M_\odot$/yr]} & 
~ & ~ & \scriptsize{[Gyr]} & \scriptsize{[$M_\odot$]} & \scriptsize{[$M_\odot$/yr]} \\
\hline\noalign{\smallskip}
\scriptsize{SFH-L1} & 30\% & $0.2_{-0.2}^{+0.3}$ & $8.6^{+0.9}_{-0.3}$ & $0.3^{+1.7}_{-0.2}$ & \scriptsize{SFH-L1}  & 35\% & $0.0^{+0.2}_{-0.0}$ & $8.6^{+0.5}_{-0.3}$ & $0.6^{+1.8}_{-0.4}$ & \scriptsize{SFH-L1}  & 35\% & $0.0_{-0.0}^{+0.2}$ & $8.6^{+0.5}_{-0.3}$ & $0.6^{+1.8}_{-0.4}$ \\
\scriptsize{SFH-L2} & 59\% & $0.8^{+0.3}_{-0.2}$ & $8.6^{+0.6}_{-0.3}$ & $0.5^{+1.8}_{-0.3}$ & \scriptsize{SFH-L2} & 33\% & $0.5^{+0.2}_{-0.1}$ & $8.6^{+0.7}_{-0.3}$ & $0.4^{+1.9}_{-0.2}$ & \scriptsize{SFH-L2} & 33\% & $0.5_{-0.1}^{+0.2}$ & $8.5^{+0.3}_{-0.2}$ & $0.2^{+0.3}_{-0.1}$ \\
\scriptsize{SFH-L3} & 11\% & $1.6_{-0.3}^{+0.3}$ & $8.5^{+0.3}_{-0.2}$ & $0.2^{+0.3}_{-0.1}$ & \scriptsize{SFH-L3} & 21\% & $0.9^{+0.2}_{-0.2}$ & $8.5^{+0.9}_{-0.3}$ & $0.3^{+1.5}_{-1.1}$ & \scriptsize{SFH-L3} & 14\% & $0.8_{-0.0}^{+0.2}$ & $8.6^{+0.9}_{-0.3}$ & $0.4^{+1.9}_{-0.2}$ \\
-- & -- & -- & -- & -- & \scriptsize{SFH-L4} & 11\% & $1.6^{+0.3}_{-0.3}$ & $8.5^{+0.3}_{-0.2}$ & $0.2^{+0.3}_{-0.1}$ & \scriptsize{SFH-L4} & 13\% & $1.2_{-0.1}^{+0.2}$ & $8.5^{+0.6}_{-0.3}$ & $0.3^{+0.8}_{-0.1}$ \\
-- & -- & -- & -- & -- & -- & -- & -- & -- & -- & \scriptsize{SFH-L5} & 6\% & $1.8_{-0.1}^{+0.3}$ & $8.4^{+0.3}_{-0.2}$ & $0.3^{+1.9}_{-0.2}$\\
\hline
\noalign{}
\end{tabular}
\end{table*}

\section{Discussion}\label{sec:disss}

According to internal statistical metrics, i.e., the Silhouette Score (see Table 1) and class stability (differences between $t_{\mathrm{peak}}$), the optimal clustering model corresponds to $k = 4$.
It describes a consistent clustering structure due to the similarity of galaxies within each class and their adequate separation between different classes (see Fig. \ref{viol-tpeak}), given the nature of the data.
%Regarding some physical properties of the $k=4$ cluster studied by the simulation for each of the four classes, 
This scenario,  cluster $k=4$, clearly separates populations with SFR peaks that are spaced significantly apart in time: 
SFH-L1 $\sim 0.0^{+0.2}_{-0.0} \, \mathrm{Gyr}$, 
SFH-L2 $\sim  0.5^{+0.2}_{-0.1} \, \mathrm{Gyr}$,
SFH-L3 $\sim  0.9^{+0.2}_{-0.2} \, \mathrm{Gyr}$, and SFH-L4 $\sim 1.6^{+0.3}_{-0.3} \, \mathrm{Gyr}$, respectively. It suggests that there are at least four distinct evolutionary scenarios in the SFH of LAEs at $z \sim 2.0$: one class with very recent and high star-formation (SFH-L1), two classes (SFH-L2 and SFH-L3) with SFR peaks at intermediate times, and one class (SFH-L4) with older and more extended activity across cosmic time. 
SFH-L1 and SFH-L2 contain most of the galaxies, reaching 35$\%$ and 33$\%$ of the total sample, respectively, while the less numerous classes are SFH-L3 and SFH-L4, which encompass 21$\%$ and 11$\%$ of the LAEs, respectively.
Regarding the median stellar mass, we observe that the four classes have a similar median stellar mass,  $\mathrm{10^{8.5-8.6}\ M_\odot}$, while the distribution is broader for the classes SFH-L3 and SFH-L4 (see \textbf{Fig.} \ref{viol-tpeak} (b)). 
SFH-L1 and SFH-L2 have a larger proportion ($>$20$\%$) of galaxies with stellar masses above $\mathrm{10^{9.0}\,M_\odot}$, while SFH-L3 and SFH-L4 present a lower proportion but a diversity in terms of stellar mass, ranging from $\sim 10^{8.0}\, \mathrm{M_\odot}$ to $10^{11.5}\,\mathrm{M_\odot}$.
%Then, for SFH-L3, with 21\% of the LAEs, its mean is located at $10^{8.5}\,[M_\odot]$, and finally, for SFH-L4, its mean is located at $10^{8.5}\,[M_\odot]$. 

On the other hand, the separation between classes is not clear for $k=5$ because the classes SFH-L4 and SFH-L5 are similar in terms of their $t_{\mathrm{peak}}$; specifically, at $\pm 1\sigma$, both classes are the same, i.e., their $t_{\mathrm{peak}}$ overlap.
For $k=3$, none of the classes exhibits a peak of star-formation activity at the time of observation $t_{\mathrm{peak}}=0$. The one peaking closer to the moment of observation $t_{\mathrm{peak}}=0.2$ has a broader dispersion, ranging from 0 to 0.6 $\mathrm{Gyr}$, and it contains 59\% of the total LAE sample analyzed. It does not resemble the results of \cite{firestone2025odinstarformationhistories} based on observational data.

\begin{center}
%\includegraphics[width=0.8\columnwidth]{MS-curve_fit-por-clase (1).png}
%\captionof{figure}{The x-axis shows $\log_{10}(M_\star/M_\odot)$ and the y-axis $\log_{10}(\mathrm{SFR})$ [$M_\odot\,\mathrm{yr^{-1}}$]. Gray points represent individual galaxies, with density indicated by the gray scale. The black dashed line marks the empirical main sequence from \citet{2023MNRAS.519.1526P}.}
\label{Figura-MSC}
\end{center}

\section{Conclusions}

The combined analysis of the relative $\mathrm{LBT(z)}$ (equation~\ref{eq:LBT_general}) and the normalized SFR (equation~\ref{eq:SFR_normalized}) allows for a homogeneous representation of the temporal evolution of LAE galaxies. When plotting $\mathrm{SFR_{n,i}}$ as a function of $\mathrm{LBT_i}$, most galaxies show a single, well-defined peak in SF activity, which is typical of LAEs. This behavior characterizes LAEs as galaxies that experience a dominant burst of SF, followed by a gradual decline in activity. This pattern provides a solid framework for studying the processes of stellar growth, cooling, and possible rejuvenation throughout their cosmological evolution. The k=4 clustering suggests that there are two other types of SFHs, finding at least two peaks of SF bursts and SFHs that are more extensive over time, from the moment of galaxy formation to the present day. \\
Comparing the classifications with \(k=3\) and \(k=5\) to our control (\(k=4\)), we found that the \(k=3\) case fails to preserve the correlation of key physical properties between classes. Specifically, its intermediate classes group populations with distinct SFHs, indicating a classification that is too general and lacks sufficient discriminatory power.
%By comparing the classifications between $k = 3$ and $k = 5$ classes with our control sample $(k = 4)$, we observe that the case with $k = 3$ does not adequately preserve the correlation between classes in terms of key physical properties; in particular, the intermediate classes tend to group populations with different SFHs, suggesting an overly general classification with less discriminatory power.
On the other hand, the model with $k = 5$ shows a partial correspondence with the case of $k = 4$. Although it maintains the temporal progression in the parameter $t_{\mathrm{peak}}$ and allows for a finer subdivision of the intermediate classes, it introduces a class that contains less than 10\% of the total galaxies, which compromises its statistical robustness and could reflect an overfitting of the clustering model. For these reasons, we consider that clustering with $k = 4$ is the most physically interpretable and robust method for describing the diversity in the SFHs of LAE galaxies. This model offers an optimal balance between temporal coherence, physical consistency, and statistical stability among classes, allowing us to clearly identify subpopulations with differentiated evolutionary trajectories.

Finally, our placement of LAEs and their classes in the SFR and stellar-mass plane (main-sequence) reveals that the mean relationships for SFH–L1 to SFH–L4 lie above the empirical main-sequence (MS) compiled by \citet{2023MNRAS.519.1526P}.
This systematic offset implies that, within the IllustrisTNG sample studied here, LAEs tend to be more actively forming stars than the average observed star-forming galaxies at similar stellar masses and redshifts.
This enhancement can be explained by selection effects (LAEs preferentially trace galaxies with strong recent SF and low dust attenuation), differences in SFR-mass calibrations between simulations and observations, or genuine physical processes (e.g., sustained cold gas inflows or merger-driven bursts).

The diversity of secondary or additional peaks observed in Fig. \ref{viol-tpeak} motivates performing a more complex classification/clustering as future work, i.e., considering  the identification of other peaks and features in the SFHs. Another relevant issue is the comparison of mock-observed SFRs, exploring different dust corrections to compare with state-of-the art observational results.

%Al comparar la clasificación entre $k = 3$ y $k = 5$ clases con nuestra muestra de control $(k = 4)$, se observa que el caso con $k = 3$ no preserva adecuadamente la correlación entre clases en términos de propiedades físicas clave; en particular, las clases intermedias tienden a agrupar poblaciones con historias de formación estelar distintas, lo que sugiere una clasificación demasiado general y con menor capacidad discriminatoria.
%Por otro lado, el modelo con $k = 5$ presenta una correspondencia parcial con el caso de $k = 4$. Si bien mantiene la progresión temporal en el parámetro $t_{peak}$ y permite una subdivisión más fina de las clases intermedias, introduce una clase que contiene menos del 10 \% del total de galaxias, lo cual compromete su robustez estadística y podría reflejar un sobreajuste del modelo de agrupamiento. Por estas razones, consideramos que el agrupamiento con $k = 4$ es el más físicamente interpretable y robusto para describir la diversidad en las historias de formación estelar de las galaxias LAEs. Este modelo ofrece un balance óptimo entre coherencia temporal, consistencia física y estabilidad estadística entre clases, permitiendo identificar con claridad subpoblaciones con trayectorias evolutivas diferenciadas.

\begin{acknowledgement}
IL acknowledges support from Proyecto Puente CIPPTE202501 (UCEN 2025). PTI acknowledges the use of GÜINA, funded by EQM200216 and ANID Vinculación Internacional 240098.
\end{acknowledgement}

%%%%%%%%%%%%%%%%%%%%%%%%%%%%%%%%%%%%%%%%%%%%%%%%%%%%%%%%%%%%%%%%%%%%%%%%%%%%%%
%  ******************* Bibliografía / Bibliography ************************  %
%                                                                            %
%  -Ver en la sección 3 "Bibliografía" para mas información.                 %
%  -Debe usarse BIBTEX.                                                      %
%  -NO MODIFIQUE las líneas de la bibliografía, salvo el nombre del archivo  %
%   BIBTEX con la lista de citas (sin la extensión .BIB).                    %
%                                                                            %
%  -BIBTEX must be used.                                                     %
%  -Please DO NOT modify the following lines, except the name of the BIBTEX  %
%  file (without the .BIB extension).                                       %
%%%%%%%%%%%%%%%%%%%%%%%%%%%%%%%%%%%%%%%%%%%%%%%%%%%%%%%%%%%%%%%%%%%%%%%%%%%%%% 

\bibliographystyle{baaa}
\small
\bibliography{bibliografia}
 
\end{document}